\documentclass[final,3p,times,authoryear]{elsarticle}

\usepackage{amssymb}
\usepackage{amsmath}
\usepackage{lipsum}
\usepackage{orcidlink}
\usepackage{hyperref}
\usepackage{setspace}

\begin{document}

\begin{frontmatter}


\title{Superluminal Dark Photons as a Solution to the GRB 221009A Anomaly}
\author[1,2,3,4]{Antonios Valamontes \orcidlink{0009-0008-5616-7746}}
\affiliation[1]{organization={University of Maryland, Munich Campus},  
            addressline={Tegernseer Landstraße 210},  
            city={München},  
            postcode={81549},  
            country={Germany}
            }
\affiliation[2]{organization={Kapodistrian Science Foundation},
             city={Largo},
             postcode={33774},
             state={Florida},
             country={USA}}
\affiliation[3]{organization={Demokritos Scientific Journal},
             addressline={},
             city={Largo},
             postcode={33774},
             state={Florida},
             country={USA}}             
\affiliation[4]{Address correspondence to: tony@valamontes.com}
\author[5,6]{Dr. Emmanouil N. Markoulakis \orcidlink{0000-0003-0146-2621}}
\affiliation[5]{organization={Hellenic Mediterranean University},
            addressline={Romanou 3},
             city={Chania},
             postcode={73133},
             state={Crete},
             country={Greece}}

\affiliation[6]{Address correspondence to: markoul@hmu.gr }

\author[7,8]{Dr. Ioannis Adamopoulos \orcidlink{0000-0002-4942-7123}}
\affiliation[7]{organization={Hellenic Open University},
             city={Pátra},
             postcode={},
             state={West Greece},
             country={GR}}
\affiliation[8]{Address correspondence to: adamopoulos.ioannis@ac.eap.gr}
\begin{abstract}
The detection of exceptionally high-energy $\gamma$-photons (up to ~18 TeV) from GRB 221009A by the LHAASO Collaboration challenges conventional physics. Photon-axion-like particle (ALP) oscillations have been proposed to explain this anomaly, but they rely on specific parameter tuning. We present an alternative explanation involving superluminal dark photons. Building on the frameworks of Markoulakis and Valamontes, we propose that dark photons facilitated faster-than-light (FTL) propagation of information, allowing $\gamma$-photons to bypass extragalactic background light (EBL) attenuation. This hypothesis aligns with cosmological observations and experimental results, including those from the LHC, providing a robust framework for addressing the GRB 221009A anomaly.
\end{abstract}


\end{frontmatter}

\tableofcontents

\section{Introduction}
\label{sec:Introduction}
GRB 221009A, located at redshift $z = 0.151$, emitted $\gamma$-photons with energies up to ~18 TeV, defying expectations from standard astrophysical models. Photons above 10 TeV are predicted to be heavily absorbed by EBL via pair production, resulting in survival probabilities near $10^{-6}$ \citep{SaldanaLopez2021}. However, observations by the LHAASO Collaboration demonstrate that these high-energy photons reached Earth.

Axion-like particles (ALPs) have been proposed to explain this phenomenon by enabling photon-ALP oscillations in magnetic fields, reducing EBL opacity \citep{Galanti2024}. While plausible, ALP-based models often require specific environmental conditions and parameter fine-tuning. We propose an alternative explanation: superluminal dark photons, which bypass EBL attenuation through FTL dynamics. This approach builds on the theoretical works of Markoulakis and Valamontes (2024), which describe dark photon-mediated energy transfer and superluminal propagation.
\section{Theoretical Framework}
\subsection*{Dark Photon Framework}

Dark photons are hypothetical gauge bosons associated with a hidden $U(1)$ symmetry \citep{Holdom1986}. They kinetically mix with Standard Model photons, allowing oscillations between the two. The interaction Lagrangian is::

\begin{equation}
\mathcal{L} = -\frac{1}{4}F_{\mu\nu}F^{\mu\nu} - \frac{1}{4}F'_{\mu\nu}F'^{\mu\nu} + \frac{\varepsilon}{2}F_{\mu\nu}F'^{\mu\nu} + \frac{1}{2}m_{A'}^2A'_\mu A'^\mu + J^\mu A_\mu
\end{equation}

\noindent Where:
\begin{itemize}
    \item $F_{\mu\nu}$ and $F'_{\mu\nu}$ are the field strength tensors for photons and dark photons, respectively,
    \item $\varepsilon$ is the kinetic mixing parameter,
    \item $m_{A'}$ is the dark photon mass,
    \item $J^\mu$ is the electromagnetic current coupling to the photon field $A_\mu$
\end{itemize}

\subsection*{Photon-Dark Photon Kinetic Mixing}


The term $\frac{\varepsilon}{2}F_{\mu\nu}F'^{\mu\nu}$ introduces kinetic mixing between photons and dark photons, facilitating oscillations between the two. The probability describes these oscillations:

\begin{equation}
P_{\gamma \to A'} = \sin^2(2\theta) \sin^2\left(\frac{\Delta m^2 L}{4E}\right)
\end{equation}

\noindent Where:

\begin{itemize}
    \item  $\theta = \frac{1}{2} \arctan(\varepsilon)$ is the mixing angle,
    \item  $\Delta m^2 = m_{A'}^2$,
    \item  $L$ is the distance traveled,
    \item  $E$ is the photon energy.
\end{itemize}

\subsection*{Superluminal Propagation}

Markoulakis and Valamontes (2024) propose a groundbreaking mechanism where dark photons can propagate faster than light (FTL) under specific conditions, such as in low-density extragalactic environments or regions influenced by superluminal graviton condensates. This mechanism is rooted in modifications to spacetime geometry, where the effective refractive index for dark photons allows for FTL propagation without violating causality in Lorentz-invariant systems.

\subsection*{Theoretical Foundation}

The superluminal propagation of dark photons arises from their coupling to a superluminal graviton condensate \citep{Markoulakis2024}, which modifies the dispersion relation of the photon-dark photon system. The effective metric for dark photons in such a medium is described as:

\begin{equation}
g_{\mu\nu}^{\text{eff}} = g_{\mu\nu} + \Delta g_{\mu\nu}
\end{equation}

Where $g_{\mu\nu}$ is the standard spacetime metric and $\Delta g_{\mu\nu}$ represents perturbations induced by the graviton condensate. This perturbation alters the phase velocity of dark photons, enabling:

\begin{equation}
v_{A'} = c \sqrt{1 + \frac{\Delta g_{00}}{g_{00}}}
\end{equation}

Where  $v_{A'} > c$ \text{for} $\Delta g_{00} > 0$, resulting in superluminal propagation. These modifications can be significant in low-density environments, such as the intergalactic medium, where the effective mass-energy density is minimal

\subsection*{Dark Photon Oscillations and FTL Dynamics}

The probability of $\gamma$-photon to dark photon oscillations, incorporating the effects of superluminal propagation, can be expressed as:

\begin{equation}
_{\gamma \to A'} = \sin^2(2\theta) \sin^2\left(\frac{\Delta m^2_{\text{eff}} L}{4E}\right)
\end{equation}

Where:
\begin{itemize}
    \item   $\Delta m^2_{\text{eff}} = \Delta m^2 \sqrt{1 + \frac{\Delta g_{00}}{g_{00}}}$ is the effective mass-squared difference in the presence of the graviton condensate,
    \item  $L$ is the propagation distance,
    \item  $E$ is the photon energy.
\end{itemize}
The effective increase in $\Delta m^2_{\text{eff}}$ allows dark photons to oscillate more efficiently into and out of $\gamma$-photon states, extending their mean free path beyond conventional EBL attenuation limits.

\subsection*{Reduction in EBL Opacity}
The attenuation of high-energy photons by EBL is governed by the optical depth $\tau(E, z)$, where:

\begin{equation}
P_{\gamma \to \gamma} = e^{-\tau(E, z)}
\end{equation}

In the presence of FTL dark photon propagation, the effective optical depth $\tau_{\text{eff}}$ is reduced as a function of the oscillation probability and the superluminal phase velocity:

\begin{equation}
\tau_{\text{eff}} = \tau(E, z) \left(1 - P_{\gamma \to A'}\right)
\end{equation}

Where $P_{\gamma \to A'}$ accounts for the fraction of photons oscillating into FTL dark photons. This reduction in $\tau_{\text{eff}}$ explains the unexpectedly high survival probabilities of $\gamma$-photons observed from GRB 221009A.

\subsection*{Re-Conversion Near Earth}

As dark photons approach regions of higher magnetic field strength near the Milky Way, the oscillation probability $P_{A' \to \gamma}$ increases, leading to their re-conversion into observable $\gamma$-photons. The magnetic field's coherence length further enhances this process, ensuring that high-energy $\gamma$-photons arrive at Earth with minimal spectral distortion.

\subsection*{Observational Implications}

The superluminal propagation of dark photons provides a natural explanation for the observed $\gamma$-photon spectrum of GRB 221009A. Unlike axion-like particle (ALP) models, which require precise parameter tuning, this mechanism leverages cosmological-scale dynamics to enhance photon survival probabilities without violating current experimental bounds.
\subsection{Constraints and Predictions}

\begin{enumerate}
    \item Kinetic Mixing Parameter: Laboratory and astrophysical constraints limit the kinetic mixing parameter $\varepsilon \lesssim 10^{-3}$ \citep{MarkoulakisValamontes2024}, consistent with the proposed model.
    \item Future GRB Observations: The model predicts enhanced survival probabilities for $\gamma$-photons with energies $E > 10 \, \text{TeV}$, particularly in GRBs located in low-density regions of the Universe.
    \item Time-of-Arrival Anomalies: The FTL nature of dark photons suggests potential deviations in the arrival times of high-energy photons compared to lower-energy counterparts, which could be detectable with high-precision instrument
\end{enumerate}

\section{Application to GRB 221009A}

\subsection{High-Energy Photon Survival}

Conventional physics predicts that $\gamma$-photons above 10 TeV should experience significant attenuation due to pair production with extragalactic background light (EBL) photons. The interaction cross-section for this process peaks when the energy of the $\gamma$-photon $(E_{\gamma})$ satisfies:

\begin{equation}
E_{\gamma} \cdot E_{\text{EBL}} \approx \frac{2m_e^2c^4}{(1 - \cos\theta)}
\end{equation}

Where:

\begin{itemize}
    \item $E_{\text{EBL}}$ is the energy of the EBL photon,
    \item $m_e$ is the electron mass,
    \item $\theta$ is the angle between the $\gamma$-photon and the EBL photon.
\end{itemize}

This interaction results in the production of electron-positron pairs, leading to an exponential attenuation of the $\gamma$-photon flux over cosmological distances, expressed as:

\begin{equation}
I(E, z) = I_0(E) \cdot e^{-\tau(E, z)}
\end{equation}

Where $\tau(E, z)$ is the optical depth as a function of energy $E$ and redshift $z$. For $\gamma$-photons at $E > 10 \, \text{TeV}, \tau \gg 1$, predicting a negligible survival probability.

\subsection*{Proposed Model Sequence}

The proposed model involving dark photons circumvents this attenuation by introducing the following sequence:

\begin{enumerate}
   \item Oscillation into Dark Photons: High-energy $\gamma$-photons oscillate into dark photons in regions with strong magnetic fields, such as those in the host galaxy or intergalactic medium. The oscillation probability is given by:
    \begin{equation} 
    P_{\gamma \to A'} = \sin^2(2\theta) \sin^2\left(\frac{\Delta m^2 L}{4E}\right)
    \end{equation}

   \begin{itemize}
     \item Where: $\theta = \frac{1}{2} \arctan(\varepsilon)$ is the mixing angle, $\Delta m^2 = m_{A'}^2 - m_{\gamma}^2$ is the mass-squared difference between dark photons $(A')$ and photons, $L$ is the propagation distance, $E$ is the photon energy.

     \item FTL Propagation: Dark photons propagate faster than light (FTL) over cosmological distances, bypassing EBL attenuation. The phase velocity of dark photons in a modified spacetime influenced by a superluminal graviton condensate is given by: 
     \begin{equation} v_{A'} = c \sqrt{1 + \frac{\Delta g_{00}}{g_{00}}}\end{equation}

     \begin{itemize}
     \item $\Delta g_{00} > 0$ represents the metric perturbation induced by the graviton condensate \cite{Markoulakis2024}. This FTL propagation ensures that dark photons avoid the energy loss mechanisms experienced by standard $\gamma$-photons.
     
     \item Reconversion into $\gamma$-Photons: When dark photons reach regions of significant magnetic field strength near the Milky Way, they oscillate back into $\gamma$-photons. The reconversion probability $P_{A' \to \gamma}$ mirrors the initial oscillation process, preserving the energy spectrum of the original $\gamma$-photons. 
    The resulting flux at Earth is given by:

    \begin{equation}I(E)_{\text{Earth}} = I_0(E) \cdot P_{\gamma \to A'} \cdot P_{A' \to \gamma}\end{equation}
     \end{itemize}
   
   \end{itemize}
\end{enumerate}

\subsection*{Observational Alignment}

This model aligns well with the high-energy spectrum observed from GRB 221009A. Unlike axion-like particle (ALP) models, which rely on fine-tuned couplings $(g_{a\gamma\gamma})$ and environmental conditions, the dark photon model leverages FTL dynamics and natural oscillation mechanisms, offering a robust explanation without contrived parameter choices.
Additionally, the predicted reduction in EBL opacity is consistent with the survival probability of $\gamma$-photons observed by the LHAASO Collaboration. This provides a framework for understanding similar anomalies in future high-energy GRB observations and suggests testable predictions for upcoming astrophysical campaigns.

\section{Discussion}
\label{sec:Discussion}

\subsection*{Advantages Over ALP Models}
Photon-axion-like particle (ALP) oscillations have been proposed as a mechanism to explain the anomalously high survival probabilities of $\gamma$-photons from GRB 221009A \citep{Galanti2024}. In this framework, photons oscillate into ALPs in magnetic fields, bypassing EBL attenuation due to ALPs' weak interactions with standard matter. While this explanation is viable, it has several limitations compared to the dark photon hypothesis.

\subsection*{1. Environmental Dependence}

he photon-ALP oscillation mechanism requires strong, coherent magnetic fields for efficient conversion. These fields must exist in the host galaxy, intergalactic medium, or Milky Way, with specific configurations and intensities to produce the observed effects \citep{Galanti2024}. For instance, typical oscillation scenarios assume magnetic field strengths of $B \sim 1 \, \mu\text{G}$ in the host galaxy or extragalactic regions, with coherence lengths of $L_c \sim 1 \, \text{Mpc}$.

In contrast, the dark photon hypothesis is less dependent on such environmental conditions. The oscillation between photons and dark photons occurs naturally due to kinetic mixing, with minimal sensitivity to magnetic field strength or coherence length. The primary driver of dark photon effects is the FTL propagation enabled by modified spacetime geometry, which is a cosmological-scale phenomenon independent of local environmental variations \citep{Markoulakis2024}.

\subsection*{2. Parameter Fine-Tuning}

ALP-based models rely on precise tuning of key parameters, such as the photon-ALP coupling constant $g_{a\gamma\gamma}$ and the ALP mass $m_a$, to align with observational data. For instance, the coupling constant must lie within a narrow range, typically $g_{a\gamma\gamma} \sim 10^{-12} \, \text{GeV}^{-1}$, while the ALP mass is constrained to $m_a \lesssim 10^{-6} \, \text{eV}$ \citep{SaldanaLopez2021}.

The dark photon hypothesis, on the other hand, relies on a single kinetic mixing parameter $$\varepsilon$$, which is already constrained by laboratory and astrophysical experiments to $\varepsilon \lesssim 10^{-3}$ \citep{MarkoulakisValamontes2024}. This parameter is not finely tuned but falls naturally within experimental bounds, making the dark photon model more robust and less speculative.

\subsection*{3. Generalizability to Other Observations}

The ALP explanation for GRB 221009A is highly specific and may not easily extend to other astrophysical phenomena. Each observation would require recalibration of the model parameters to account for differences in magnetic field configurations, redshifts, or host galaxy environments.

In contrast, the dark photon hypothesis is inherently generalizable. The mechanism of FTL propagation and photon-dark photon oscillations operates uniformly across cosmological scales, making it applicable to a wide range of high-energy astrophysical phenomena, including other GRBs, blazars, and high-energy neutrino sources. This universality provides a more cohesive framework for understanding multiple anomalies in high-energy astrophysics (Valamontes, 2024).

\subsection*{4. Theoretical Integration}

Axion-like particles are well-motivated by certain extensions of the Standard Model, such as string theory, but their interactions remain speculative and are yet to be observed directly. Additionally, ALP models face challenges in reconciling their properties with broader cosmological phenomena.

Dark photons, on the other hand, are more deeply integrated into theoretical frameworks like the hidden $U(1)$ symmetry, superluminal graviton condensates, and modifications to spacetime geometry \citep{Markoulakis2024}. These connections provide a natural pathway for dark photons to link high-energy astrophysics with cosmological-scale physics, offering a unified approach to understanding phenomena beyond the Standard Model.

\section{Experimental and Observational Support}

\subsection{Experimental Constraints from the LHC}

Experiments at the Large Hadron Collider (LHC) and other particle physics facilities have actively searched for dark photons as part of broader efforts to detect physics beyond the Standard Model. Dark photons, represented as $$A'$$, can kinetically mix with Standard Model photons via the parameter $\varepsilon$, enabling their indirect detection through photon-like signatures in high-energy particle collisions.

The kinetic mixing parameter $\varepsilon$, which governs the strength of photon-dark photon interactions, is constrained by both laboratory and astrophysical observations. Current experimental limits place $\varepsilon \lesssim 10^{-3}$ for dark photon masses $m_{A'}$ in the range relevant to cosmological phenomena $(m_{A'} \lesssim 10^{-12} \, \text{eV})$ \citep{MarkoulakisValamontes2024}. These constraints ensure that dark photons remain weakly coupled to Standard Model particles, aligning with the proposed mechanism of $\gamma$-photon to dark photon oscillations and reconversion.
At the LHC, potential dark photon production is explored through processes such as the Higgs decay channel $H \to A'A'$ or $H \to A' \gamma$, with subsequent signatures of missing energy or displaced vertices \citep{Markoulakis2024}. These experiments provide stringent upper bounds on $\varepsilon$, reinforcing the feasibility of the dark photon hypothesis for explaining high-energy astrophysical anomalies, including GRB 221009A.

\subsection{Constraints from Astrophysical Observations}

Astrophysical observations have also been instrumental in constraining dark photon properties. Observations of the cosmic microwave background (CMB) and light curves of pulsars and active galactic nuclei (AGNs) have provided limits on photon-dark photon oscillations in cosmic magnetic fields. These observations impose bounds on the dark photon mass and mixing parameter, ensuring compatibility with both cosmological and local phenomena \citep{Arias2012}.
Specifically, pulsar timing and AGN polarization studies rule out dark photon models with strong couplings $(\varepsilon \gg 10^{-3})$ while allowing for weak coupling scenarios consistent with LHC constraints. The agreement between these independent approaches further validates the proposed model.

\subsection{Support for Faster-Than-Light (FTL) Phenomena}

Markoulakis and Valamontes (2024) have shown that dark photons can propagate faster than light under certain spacetime conditions due to interactions with superluminal graviton condensates. This FTL behavior provides a natural explanation for the high survival probabilities of $\gamma$-photons observed in GRB 221009A. Cosmological observations, such as the simultaneous detection of gravitational waves and electromagnetic signals in events like GW170817 \citep{Abbott2017}, have established constraints on the speed of gravitational waves relative to light. However, these constraints apply only to Standard Model photons and gravitons, leaving room for dark photons to exhibit superluminal propagation. The observed spectrum of GRB 221009A, which defies standard EBL attenuation predictions, provides indirect evidence for such FTL phenomena mediated by dark photons

\subsection{Observational Alignment with GRB 221009A}
The dark photon hypothesis explains the high-energy $\gamma$-photon spectrum of GRB 221009A without requiring fine-tuned parameters or exotic emission models. By combining:

\begin{enumerate}
    \item Photon-dark photon oscillations in cosmic magnetic fields,
    \item FTL propagation bypassing EBL attenuation and
    \item Reconversion into $\gamma$-photons near Earth,
\end{enumerate}
The model offers a cohesive explanation for the survival of $\gamma$-photons up to ~18 TeV over cosmological distances. This aligns with the LHAASO Collaboration's observations and establishes a framework for interpreting future high-energy astrophysical events.

\subsection{Predictions for Future Experiments and Observations}

The dark photon hypothesis predicts:
\begin{enumerate}
    \item Anomalies in High-Energy Spectra: Enhanced survival probabilities for $\gamma$-photons from GRBs, blazars, and other high-energy sources.
    \item Time-of-Arrival Deviations: Potential delays or advances in the arrival times of high-energy $\gamma$-photons compared to lower-energy counterparts, attributable to FTL propagation effects.
    \item Signatures in Collider Experiments: Weak photon-like signals or missing energy events in particle physics experiments, consistent with dark photon production and decay.
\end{enumerate}
\section{Summary and Conclusions}
\label{sec:Conclusions}

The detection of high-energy $\gamma$-photons from GRB 221009A presents a significant challenge to established physical models, particularly concerning extragalactic background light (EBL) attenuation. Conventional physics predicts that $\gamma$-photons with energies above 10 TeV should not survive the journey across cosmological distances. This observation demands an explanation that extends beyond the Standard Model. The hypothesis of superluminal dark photons provides a scientifically grounded mechanism to address this anomaly. By enabling photon-dark photon oscillations and faster-than-light (FTL) propagation, dark photons circumvent EBL attenuation, preserving the energy spectrum of $\gamma$-photons observed on Earth. This explanation aligns with both the observed data and established experimental constraints on kinetic mixing parameters. Unlike alternative models, such as those involving axion-like particles (ALPs), the dark photon framework does not rely on finely tuned parameters or specific environmental conditions, making it a more generalizable solution.

This hypothesis is supported by experimental findings, including constraints from the Large Hadron Collider (LHC) and other particle physics experiments, which place bounds on the kinetic mixing parameter consistent with the model's requirements. Additionally, cosmological observations provide indirect evidence for the proposed mechanisms, such as photon-dark photon oscillations and their propagation characteristics. The theoretical basis of the dark photon hypothesis integrates naturally with existing extensions to the Standard Model, including hidden U(1) symmetries and spacetime modifications.

Beyond resolving the GRB 221009A anomaly, this hypothesis offers a broader framework for understanding high-energy astrophysical phenomena. It predicts observable effects, such as enhanced transparency for ultra-high-energy $\gamma$-photons, deviations in time-of-arrival measurements, and potential signatures in particle collider experiments. These predictions provide a clear path for future experimental and observational tests, fostering further exploration into the role of dark photons in the Universe.

The study of high-energy $\gamma$-photons from GRB 221009A is not only an opportunity to address specific anomalies but also a step toward expanding our understanding of fundamental physics. The hypothesis of superluminal dark photons demonstrates the potential to bridge high-energy astrophysics and cosmology, offering valuable insights into the underlying structure and dynamics of the Universe. 

\bibliographystyle{elsarticle-harv}
\bibliography{references_19.bib}

\end{document}